# BAHULAM: *Distributed Data Analytics on Secure Enclaves*

*Srinivasa Rao Aravilli, Cisco Systems Inc*

## Abstract

This is a survey of some of the currently available frameworks (opensource/commercial) in order to run distributed data applications (Apache Hadoop, Spark) on secure enclaves. Intel, AMD, Amazon support secure enclaves on their systems Intel-SGX, AMD Memory Encryption, AWS Nitro Enclaves respectively.

Keystone is an open source framework for architecting Trusted Execution Environments and isolation of software.

*Bahulam* means $10^{23}$ in Sanskrit language, which is apt for large scale data processing in secure way.

## Introduction

In the current scenario of application development and deployment, application security and data security are key for any organization specially to protect integrity and confidentiality. One way of achieving high security using encryption (homomorphic encryption) of data, but it may impact the performance to significant level and not many organizations are using fully homomorphic encryption. Secure enclaves are new innovation started in the last few years by hardware vendors as well some opensource options. Secure enclaves provide applications isolations by creating enclaves in the hardware and running the applications on the enclaves. In this way, other applications will not be able to access the data/applications which are running in the enclaves because enclaves provide trusted execution environment and remote attestation of the code as well.

Apache Spark is commonly used distributed application framework in many enterprises. The following are some of the frameworks analyzed in the survey.

- Maru SGX Spark
- Opaque: Secure Apache Spark SQL
- VC3 - Verifiable Confidential Cloud Computing
- SGX-PySpark: Secure Distributed Data Analytics
- AWS Nitro Enclave

## Maru - SGX Spark

Intel SGX (Software Guarded Extensions) provides enclave to support isolated memory regions for code and data to execute through MEE (memory encrypted engine). SGX are extensions to Intel x86-64 ISA (Instruction Set Architecture).

Maru SGX Spark developed by the team from The Imperial college of London to run Apache Spark workload on Intel SGX servers in a secure way.

SGX Spark executes only sensitive parts of the Spark inside enclave in separate JVM and code outside enclave access the encrypted data. It uses two different JVM's (one inside the enclave) and the other outside the enclave and both will execute the Spark Jobs and coordinate. The main challenges in this approach is handling spark partition, data passing among the JVMs, and to handle memory efficiently. Intel SGX provides only 94MB of memory in the enclave for the applications, so Spark Job needs to be partitioned keep this in mind.

Maru divides the spark in such way that only minimal amount of spark code moved to Enclave to run and rest will be run in the outside JVM itself. The following example which is motioned in their paper.

GitHub Location: https://github.com/lsds/sgx-spark

Maru provides a docker instance which can be run on a SGX enabled Intel processors. Maru provides few examples like word count application, line count and K-Means clustering (Machine Learning). The SGX Spark implementation leverages **sgx-lkl,** a library OS that allows to run Java-based applications inside SGX enclaves. They currently support only RDD based Spark Applications and DataFrames/DataSets support needs to be added, since it is open sourced so others will be able to add the support for Spark Data Frames and Data Sets as well.

Detailed Presentation and Talks: SGX-Spark

**Opaque:** Secure Apache Spark SQL

Opaque framework designed and developed by riselab team from US Berkley, to run Apache Spark SQL on Intel SGX and to support strong security for Spark SQL queries.

Spark shuffles data among the worker nodes to execute map reduce jobs and any insider in the organization may be able to infer the key data during this process by inspecting the data. Opaque supports oblivious mode to execute SQL queries. With Opaque one of the side channel attack i.e data access patterns can be prevented using their oblivious mode. Oblivious mode helps to hide the data access patterns at network level. Opaque supports oblivious mode and execute operations map, reduce, join, filter, aggregations but not all SQL operations of Spark. They have reimplemented Spark SQL Catalyst engine using C++ in order to execute Spark SQL queries on SGX enclaves.

They have executed Bigdata benchmarks SQL workload with encryption mode and didn't observed major performance degradation when running the code in encrypted in the enclave compared to normal. Opaque compared with GraphSC (STOA oblivious graph processing library) as well and details can be referred from the GitHub repository.

Opaque system was demoed in one of the Spark Summit and showed how Spark Queries can be the victim of side channel attacks and how it can be prevented using oblivious mode with encrypted DAG etc.

Git Hub Location:
https://github.com/ucbrise/opaque

Opaque currently supports only Spark SQL security and other Spark modules like Spark ML, GraphX and any other Spark arbitrary jobs are not yet supported.

It makes use of Intel SGX SDK and provide a docker instance with prebuilt version of opaque to experiment.

**VC3:** Verifiable Confidential Cloud Computing

VC3 is one of the first systems allows to run Hadoop Map reduce jobs in the AWS Cloud in a secure way. VC3 protects map reduce computation from untrusted cloud and ensures confidentiality of the analytics and data using trusted Intel SGX processors and compiler enforced variants. VC3 developed by Microsoft research team and relies on SGX processors to isolate memory regions and run map reduce jobs. VC3 executed using HDInsight Hadoop distribution on Windows. VC3 implements two security protocols i.e. to attest each map reduce job in the cloud and for key exchange to run the job. They have modified the C++ compiler to support integrity checks in the enclave. VC3 is achieved similar kind of performance with normal Hadoop with negligible runtime overhead.

VC3 explained in detailed in the Microsoft Research site and paper can be downloaded. VC3 Paper

A talk was given on VC3 by Microsoft team in IEEE Symposium on Security and Privacy

**AWS Nitro Enclaves**

Amazon recently launched AWS Nitro Enclaves. Nitro Enclaves supports isolated compute environments and cryptographic attestation of the software. Nitro Enclaves uses the same Nitro Hypervisor technology that provides CPU and memory isolation for EC2 instances.

As of writing this paper, Nitro Enclaves are offered only for preview customers, so not sure what kind of workloads Nitro enclaves supports and other features like performance, encryption overheads etc. Amazon Nitro Enclaves are only supported in the cloud or it can be supported in the client-side devices like Greengrass, is not known at this point of time. It will be ideal to have Nitro Enclaves on the client side as well to support strong security for the IoT devices and IoT use cases. It is not unknown whether AWS Nitro enclaves are built using the open source Keystone Framework or different one and portability across other vendors as well.

## SGX-PySpark

PySpark is developed on top of Apache Spark and provides APIs in Python to create and execute Map reduce jobs.

SGX-PySpark (secure distributed data analytics system) developed to run PySpark on Intel SGX. SCONE is a shielded execution framework to enable unmodified applications to run inside SGX enclaves. SGX-PySpark system built using SCONE which make use of compiler-based approach to run the applications in SGX enclaves. They have compiled JVM, CPython/Pypy using SCONE compiler in order to run Pyspark executors in the SGX enclaves. SGX-PySpark consists of two main components: 1) Configuration and attestation service (CAS) component and 2) PySpark with integration with the SCONE library to run inside enclaves.

GitHub Location: https://github.com/doflink/sgx-pyspark-demo

SGX-PySpark benchmarked TPC-H (decision support system benchmarks) queries and compared performance with native py-spark vs sgx-pyspark and concluded that not much performance overhead with sgx-pyspark.

## References

1) Apache Spark. https://spark.apache.org
2) Py4J. http://py4j.sourceforge.net
3) PySpark. http://spark.apache.org/docs/2.4.0/api/python/pyspark.html
4) TPC-H Benchmark. http://www.tpc.org/tpch
5) SCONE: https://sconedocs.github.io/
6) Opaque : https://github.com/ucbrise/opaque
7) MARUSGX-SPARK: https://github.com/lsds/sgx-spark
8) Keystone: An Open Framework for Architecting TEEs, arXiv:1907.10119
9) Keystone – Enclaves: https://keystone-enclave.org/
10) AWS Nitro Enclaves: https://aws.amazon.com/ec2/nitro/nitro-enclaves/
11) BIG Data Benchmarks: https://amplab.cs.berkeley.edu/benchmark/
12) SGX - https://software.intel.com/en-us/sgx
13) SGX-LKL: Securing the Host OS Interface for Trusted Execution, arXiv:1908.11143v1 [cs.OS] 29 Aug 2019
14) K. Nayak, X. S. Wang, S. Ioannidis, U. Weinsberg, N. Taft,and E. Shi. GraphSC: parallel secure computation made easy. In Proceedings of the 36th IEEE Symposium on Security and Privacy (S&P), 2015.
15) Do Le Quo, Franz Gregor, Jatinder Singh, Christof Fetzer: SGX-PySpark: Secure Distributed Data Analytics. The World Wide Web Conference, May 2019